\documentclass[journal]{IEEEtran}
\usepackage{cite}
\usepackage[pdftex]{graphicx}
\usepackage{amsmath,amssymb,amsfonts}
\usepackage{balance}
\usepackage{array}
\usepackage{wrapfig}
\usepackage{multirow}
\usepackage{tabu}
\usepackage{csquotes}
\usepackage{xcolor,soul}
\usepackage{url}
\usepackage{float}
\usepackage{algorithm}  
\usepackage{algorithmicx}  
\usepackage{algpseudocode}
\usepackage{caption}
\usepackage{subfigure}
\ifCLASSOPTIONcompsoc
    \usepackage[caption=false, font=normalsize, labelfont=sf, textfont=sf]{subfig}
\else
\usepackage[caption=false, font=footnotesize]{subfig}
\fi

\def\BibTeX{{\rm B\kern-.05em{\sc i\kern-.025em b}\kern-.08em
    T\kern-.1667em\lower.7ex\hbox{E}\kern-.125emX}}

\begin{document}

\title{OFEI: A Semi-black-box Android Adversarial Sample Attack Framework Against DLaaS} 

\author{Guangquan Xu,~\IEEEmembership{Member, ~IEEE,}
        GuoHua Xin,
        Litao Jiao, 
        Jian Liu\IEEEauthorrefmark{1},
        
        Shaoying Liu\IEEEauthorrefmark{1},~\IEEEmembership{Fellow, ~IEEE,}
        Meiqi Feng,
        and
        Xi Zheng
\thanks{S. Liu is with the School of Informatics and Data Science, Hiroshima Univesrity, Hiroshima, Japan. Email: sliu@hiroshima-u.ac.jp}
}


\maketitle


\begin{abstract} 

With the growing popularity of Android devices, Android malware is seriously threatening the safety of users. Although such threats can be detected by deep learning as a service (DLaaS), deep neural networks as the weakest part of DLaaS are often deceived by the adversarial samples elaborated by attackers. In this paper, we propose a new semi-black-box attack framework called one-feature-each-iteration (OFEI) to craft Android adversarial samples. This framework modifies as few features as possible and requires less classifier information to fool the classifier. We conduct a controlled experiment to evaluate our OFEI framework by comparing it with the benchmark methods JSMF, GenAttack and pointwise attack. The experimental results show that our OFEI has a higher misclassification rate of 98.25\%. Furthermore, OFEI can extend the traditional white-box attack methods in the image field, such as fast gradient sign method (FGSM) and DeepFool, to craft adversarial samples for Android. Finally, to enhance the security of DLaaS, we use two uncertainties of the Bayesian neural network to construct the combined uncertainty, which is used to detect adversarial samples and achieves a high detection rate of 99.28\%.

\end{abstract}


\IEEEpeerreviewmaketitle


\section{Introduction}
\IEEEPARstart{D}{eep} learning has been widely used in various tasks, such as vision, speech recognition, language processing, and financial fraud detection. With deep learning, large data sets can be used to achieve higher accuracy than was possible with previous classification techniques\cite{liu2017survey}. With the emergence of deep learning as a service (DLaaS), it is easier to apply deep learning to the processing of complex data in many fields. On the DLaaS platforms provided by Microsoft and Google, users can directly use the deep learning services. In Android malware detection, DLaaS can be used to effectively identify various variants of malware\cite{watson2015malware}\cite{hatem2014malware}.\par
Although deep learning is powerful, the deep learning algorithm is the weakest part of the DLaaS system. Because the deep learning algorithm is designed on the premise of the equivalent probability distribution of training data and testing data, attackers can carefully design attacks that do not adhere to this hypothesis to achieve the effect of deceiving the neural network \cite{naway2018review}. Recent research has shown that attackers can easily change the output of a deep neural network (DNN) by adding relatively small disturbances to the input\cite{szegedy2013intriguing}. Therefore, many effective algorithms have been proposed to construct adversarial samples \cite{goodfellow2015explaining}\cite{papernot2016limitations}\cite{qiu2020adversarial}\cite{moosavi2016deepfool} and the attacker can launch a variety of attacks on the DNN in the testing phase. This paper mainly studies evasion attacks on malware detection. Without affecting the training data, the attacker carefully interferes with a malicious instance so that it is incorrectly identified as benign by the trained model.\par
Fortunately, the attacker cannot obtain the architecture and parameters of the model in DLaaS. Because DLaaS is a black box system, attackers can only call the corresponding APIs to obtain useful information \cite{goodman2019cloud}\cite{goodman2020attacking}. Therefore, the DLaaS deployed on the cloud is more difficult to attack, which provides a sense of security to users. However, recent research has shown that attacks on cloud services can occur. Attackers can attack the model by querying the parameters and architecture of the cloud service model \cite{goodman2020transferability}, which consumes many resources in reality. Transferring learning attacks relies on a large collection of open-source models. The adversarial samples are transferred between the models trained on the same data set \cite{liu2016delving}. In addition, the attacker can also deceive the Google Cloud Vision API \cite{hosseini2017google} by performing a spatial transformation on the image.\par
A large number of attacks are aimed at image classification on DLaaS. To the best of our knowledge, there is no semi-black-box attack method to craft adversarial samples against DLaaS-based Android malware detection model. In the image field, the production of adversarial samples for DNNs usually involves adding a small amount of carefully adjusted perturbations, which are imperceptible to the human eye. The modified images are identified as completely different categories by the classifier. However, the attack algorithm used to craft adversarial samples needs to modify too many pixels in each iteration to deceive the target classifier. Different from adversarial images, there are strict restrictions on the production of Android adversarial samples, which are as follows: 1) The features of Android adversarial samples are usually not represented by continuous real numbers. For example, Android applications use or do not use a certain system call. Therefore, a binary value vector is used to represent an application. 2) The attacker must interfere with the application program without destroying the application program function to maintain the discreteness of the features. Unlike the continuous change of pixel values in the image, to ensure that the functions of the original Android package (APK) are not damaged, we can not delete the features in the adversarial sample and only add a limited number of features.\par
To ensure that Android malware classification based on DLaaS has a more secure DNN interface, we mainly study the characteristics of Android adversarial samples to further understand attacks. Finding the characteristics of the adversarial area is the key to providing a good defense. Szegedy \cite{szegedy2013intriguing} believes that an adversarial area is a place with low probability density. Therefore, Feinman \cite{feinman2017detecting} proposed the use of kernel density detection against samples. However, this density-based method often has limitations in regard to representing the adversarial areas. When the adversarial sample and the normal sample are surrounded by the same number of samples, they cannot be distinguished. Ma X \cite{ma2018characterizing} proposed local intrinsic dimensions to describe the local structure around the adversarial sample. It regards the space surrounding the adversarial sample as a subspace that spans multiple manifolds. The local intrinsic dimension of this subspace is larger than any one of these manifolds considered separately. It is different from measuring the characteristics of the adversarial space through the neighbors of the adversarial sample. We use the Bayesian uncertainty estimation to solve the situation where density estimation cannot distinguish adversarial samples. Aleatoric uncertainty (AU) is the inherent noise in the data, which may be an error generated by extracting features from malware. Epistemic uncertainty (EU) is an error of the trained model. We use these two types of uncertainties \cite{kendall2017uncertainties} to construct the combined uncertainty, which characterizes the adversarial area and distinguishes the adversarial sample from the normal sample for defense.\par
In this paper, we aim to study the attack and defense framework for Android malware classification based on DLaaS. In particular, our contributions are as follows:\par
\begin{itemize}
\item We propose a semi-black-box attack method, i.e., one-feature-each-iteration (OFEI), in the Android adversarial sample scenario. This method only needs to obtain probabilistic labels from DLaaS and perturbs one feature each iteration through the simulated annealing algorithm to deceive the DNN classifier. On the Virusshare \cite{virusshare.com} and Contagio \cite{contagiodump.blogspot.com/} datasets, the OFEI achieves similar effects to the white-box attack algorithm JSMF \cite{grosse2016adversarial}. Compared with GenAttack \cite{alzantot2019genattack} and pointwise attack \cite{li2020adversarial}, the OFEI has a higher misclassification rate and less disturbance. In addition, OFEI has fewer queries on DLaaS. Since gradient information is not required, the OFEI can attack many types of DNNs, even network structures with difficult gradient calculations. 
\item The OFEI we proposed can be used as an attack framework to convert the white-box attack algorithm that was originally not suitable for Android adversarial sample production into an attack algorithm for Android adversarial samples.
\item To make DLaaS more secure, we propose a defense mechanism against attacks on Android malicious samples. Under the Bayesian neural network, we detect the combined uncertainty differences between adversarial samples and normal samples. The detection effect of this method is better than that of local intrinsic dimension and kernel density, and the detection rate reaches 99.28\%.
\end{itemize}
The rest of this paper is organized as follows: Section 2 describes the related work about adversarial sample attacks and defense against DLaaSs. Section 3 introduces the groundwork of this paper. In Section 4, we show the proposed OFEI attack framework. Section 5 discusses the uncertainty to defend against adversarial sample attacks. Section 6 analyzes the experimental results. Finally, Section 7 concludes the paper.

\section{Related Work}
\label{section:related work}
\subsection{The Security Issues of DLaaS}
With the large-scale application of DLaaS, security issues have gradually attracted the attention of cloud service companies \cite{xu2020trust2privacy} \cite{qiu2007voltage}. The attacker obtains the predicted value by querying the API interface of DLaaS for a limited time and then infers the specific parameters and structure of the model. Shokri et al. \cite{shokri2017membership} used shadow training technology to determine whether the data record was in the training set, and then the black-box model was trained in the cloud by using the Google prediction API. Ilyas et al. \cite{ilyas2017query} proposed a black-box attack algorithm based on a natural evolution strategy, which greatly reduced the number of queries to the target model. Different from the often discussed image service, we consider the Android malware detection service deployed on the cloud. Attackers craft Android adversarial samples to mislead the service.\par
\subsection{Deep Learning Attack}
Deep learning algorithms are one of the most vulnerable parts of the DLaaS system \cite{deng2020analysis}. Because of the nonlinearity of the deep neural network \cite{szegedy2013intriguing}, a slight disturbance applied to the input by the attacker will cause the output to be changed. In addition, the linear calculation of the high-dimensional space easily enlarges the disturbance to the input \cite{goodfellow2015explaining}. Therefore, the existence of deep neural network linear part also leads to vulnerability. Furthermore, the overfitting caused by insufficient model averaging and regularization makes the attack possible.\par
In the image field, there are many adversarial sample attack methods. Goodfellow \cite{goodfellow2015explaining} proposed the fast gradient sign method (FGSM) which performs multiple iterations and adds tiny disturbances to each pixel. Papernot et al. \cite{papernot2016limitations} proposed the jacobian-based saliency map attack (JSMA) method, which constructs adversarial saliency maps through the forward derivative. The JSMA method can effectively explore the feature space of the adversarial sample and modify the feature that has the greatest impact on the output. Moosavi-Dezfooli et al. \cite{moosavi2016deepfool} proposed the DeepFool method to calculate the minimum necessary disturbance and applied it to the construction of adversarial samples. The L2 norm is used to limit the disturbance scale during the attack. The DeepFool attack on the model is better than the FGSM and JSMA methods. Alzantot et al. \cite{alzantot2019genattack} proposed the GenAttack method for the black box, which is based on a genetic algorithm to search for adversarial samples and does not need to know the gradient information of the classifier. Compared with GenAttack, our proposed OFEI greatly reduces the number of queries.\par 
In the field of Android malware detection, the attack methods are limited by the application scenarios. Grosse et al. \cite{grosse2016adversarial} applied the JSMA method from a continuous differentiable space to discretely restricted malware detection, which proves that adversarial attacks do exist in the domain of malware detection. With the JSMF attack method, the misclassification rate of the classifier reaches 80\% in the experiment. Compared with JSMF, our proposed OFEI attack framework has a higher misclassification rate. Li et al. \cite{li2020adversarial} proposed the pointwise attack, which first adds noise perturbation to the malware to construct an adversarial sample, and then modifies features to generate an adversarial sample with the least perturbation. Compared with the pointwise attack, OFEI has fewer disturbances and fewer query times.\par
\subsection{Deep Learning Defense}
A large number of methods for detecting adversarial samples and enhancing the robustness of the model have been proposed to defend against attacks\cite{li2020intelligent}\cite{zeng2020data}\cite{zhou2019generalized}. Szegedy et al. \cite{szegedy2013intriguing} proposed adversarial training. The defender adds adversarial samples to the training data to strengthen the robustness of the trained model. Dujaili et al. \cite{al2018adversarial} proposed minmax adversarial training, which extends the adversarial training to the field of malware. Minmax adversarial training generates adversarial samples from all malware in the training set and calculates their loss function value, and then minimizes the loss of adversarial samples and benign software during the training process. Li et al. \cite{li2020adversarial} proposed the adversarial deep ensemble method, which performs minmax adversarial training on each basic classifier, and then combines multiple classifiers. Adversarial training relies on the prior knowledge of adversarial samples and requires a large number of adversarial samples to train the model. Papernot et al. \cite{papernot2016distillation} proposed defensive distillation which modifies the SoftMax layer to produce an output closer to the average output and uses it to train a more robust model. Ma X et al. \cite{ma2018characterizing} proposed a detection method based on local intrinsic dimensionality and uncertainty, which uses the characteristic that the values of adversarial samples are much larger than those of normal samples to identify adversarial samples.

\section{Preliminaries}
\label{section:Flow}
In this section, we first introduce the background of the Android application and the process of extracting features in the Drebin classifier. Then we apply a simple deep neural network on DLaaS to detect malware. We summarize the commonly used symbols in TABLE \ref{tb:table0}.
\begin{table}
\begin{center}
\small
\caption{The table of commonly used symbols.}
\begin{tabular}{|c|c|}
\hline
$Z$ & A abstract space of all APK files \\
$z$ & An Android APK file \\
$X$ & A abstract feature space of Android samples \\
$x$ & A feature vector of Android sample \\
$\Phi\left (  \right )$ & Feature mapping function \\
$f(x)$ & The result vector DLaaS returns\\
$c$ & The categories of Android samples\\
$\delta(x)$ & The adversarial disturbance \\
$t$ & Number of DLaaS queries \\
$T$ & Simulated annealing temperature \\
\hline
\end{tabular}
\label{tb:table0}
\end{center}
\end{table}

\subsection{Android Background}
We use Android malware as our attack case and defense algorithm. Since it essentially solves the discrete problem of attacks, it can be extended to other types of malware. The difference is that the modifiable features are limited by the specific malware type. After the APK is decompressed, it mainly includes Android manifest.xml and classes.dex. We use these two files to extract features in the following text, so we briefly introduce the information contained in the files.\par
Android manifest.xml: The configuration file of the Android application that has been compiled, including the name, version number, activity, permission, service, and other information of the Android APK. The manifest declares multiple app components, as follows: activity, broadcast receiver, service, and content provider. An activity is usually a separate visual interface, on which some controls are displayed and can monitor or respond to user events. The broadcast receiver receives and responds to broadcasts. A service is an application component that is executed in the background for a long time without a user interface. When sharing data between different applications, the content provider provides the data of one application to other applications.\par
classes.dex: Dalvik bytecode file, which is executed by the Dalvik virtual machine. The Android program code is compiled and encapsulated in this file, which includes the API calls and the data information used by the application. The data information includes the IP address, hostname, and URL in the disassembled code. Malware usually makes calls to restricted or suspicious APIs and uses data information to establish network connections and disclose user information.\par
\subsection{Drebin}
Drebin \cite{arp2014drebin} is a lightweight method used to detect Android malware, which mainly uses static analysis. Drebin extracts features from an Android manifest.xml and classes.dex and organizes them into string sets according to 8 different feature sets, as shown in TABLE~\ref{tb:table1} .\par

\begin{table}
\small
\caption{Feature sets of Drebin.}
\begin{tabular}{llll}
\hline
\textbf{File type} & \textbf{ID} & \textbf{Name} & \textbf{Number} \\
\hline 
manifest & $S_{1}$ & Hardware components  & 44\\
& $S_{2}$ & Requested permissions & 841 \\
& $S_{3}$ & Application components & 23775\\
& $S_{4}$ & Filtered intents & 2076\\
dexcode & $S_{5}$ & Restricted API calls & 149\\
& $S_{6}$ & Used permission & 37\\
& $S_{7}$ & Suspicious API calls & 36\\
& $S_{8}$ & Network addresses & 118\\
\hline
\end{tabular}
\label{tb:table1}
\end{table}

In the following, we use the feature extraction form in Drebin and define a joint set S of 8 feature sets, where  $S=S_{1}\bigcup S_{2}\bigcup S_{3} \cdots \bigcup S_{8}$, which is used to map the application to the $\left | S\right |$ high-dimensional vector space. APK file z  $\in$ Z, which is the abstract space of all APK files, is designed. The feature vector $x=\left ( x_{1},x_{2},\cdots,x_{m}\right )^{T} \in X =\left \{0,1\right \}^{m}$ is constructed. The mapping $\phi:Z\rightarrow X$ is defined as each feature is we extracted from the application z. If the feature exists (the string exists), then the corresponding dimension is set to 1; otherwise, it is set to 0.

\subsection{A Simple Deep Neural Network}
The first layer of DNN usually inputs data, the hidden layer in the middle performs calculations, and the final layer outputs the results. Neurons are connected with different biases and weights and are finally passed to the next neuron through the activation function. The neurons between two adjacent layers are connected to each other, forming a fully connected network. In the process of training the model, the loss function is used to measure the output loss of the training samples. Generally, the bias and weights are determined by the gradient descent method and backpropagation iteration to optimize the loss function.\par
We define a simple neural network on DLaaS to offer the malware detection service. We use the feature extraction method of Drebin $\varphi:Z\rightarrow X$ to map an Android application $z \in Z$ into a feature vector $x \in X$, where Z is the abstract space of Android applications, and $X$ is the space of feature vectors. The input of the neural network is the extracted feature vector. We use a nonlinear activation function in (\ref{con:eqa1ac}), and the iterative formula for each layer is shown in  (\ref{con:eqa1dnn}), as follows:\par
\begin{equation}
j \in\left[1, n_{k}\right]: f_{k, j}(x)=\max (0, x)  \label{con:eqa1ac}.
\end{equation}
\begin{equation}
x=\sum_{i=1}^{n_{k-1}} w_{i, j} \cdot x_{i}+b_{i, j} \label{con:eqa1dnn},
\end{equation}
where k is the index for layers, and $n_{k}$ is the number of neurons in the k-th layer, and j is the index of the k-th layer neuron.

At the end of neural network, we use the SoftMax layer to obtain the probability distribution of each category. The neural network on DLaaS returns a result vector $f(x)=[f_{0}(x),f_{1}(x)]$, where $f_{0}(x)+f_{1}(x)=1$. $f_{0}(x)$ represents the probability that $x$ is benign, and $f_{1}(x)$ represents the probability that $x$ is malicious. Finally, we takes the category with a higher probability as the classification result $y= \mathop{\arg\max}_{i} f_{i}(x)$, where "0" means benign and "1" means malicious.

\section{OFEI: The Proposed Attack Framework}
\begin{figure}[htb]
\center{\includegraphics[width=1.0\linewidth]{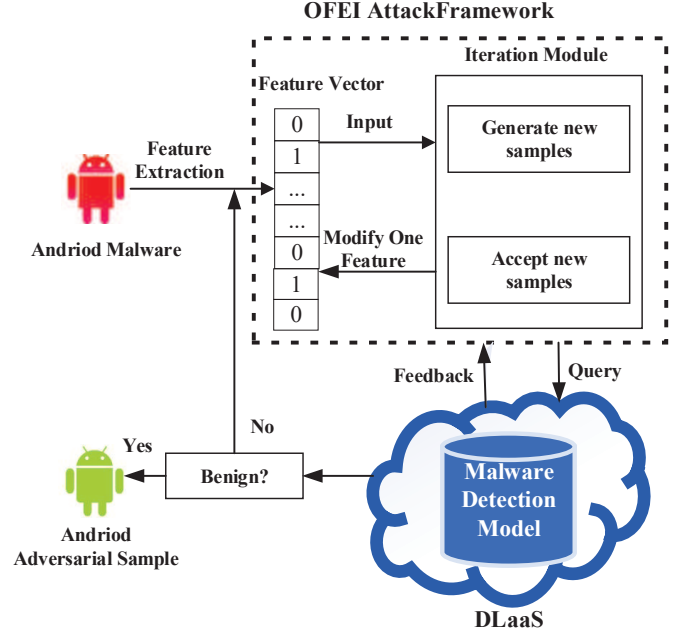}}
\caption{The architecture of our OFEI attack framework.}
\label{fig:framework}
\end{figure}

Our attack framework is shown in the Fig.~\ref{fig:framework}. The OFEI attack framework includes feature extraction and iteration modules. The attacker extracts the features of the malware and inputs them into the iterative modules. The iterative modules consist of the following two parts: new sample generation and acceptance of the new samples. The iteration modules query the DLaaS and obtain feedback which includes the probability label of the classification results. Then, the iteration modules rely on feedback to modify the features of the Android sample. The attack framework continues to iterate until the result of the classifier changes. In Section 4.1, we describe the evasion attack on Android samples. In Section 4.2, we analyze the limitations of the attack. In Section 4.3, we show the OFEI for crafting Android adversarial samples against DLaaS. In Section 4.4, we use the attack framework to extend other attack methods.
\subsection{Evasion Attack}
In the evasion attack, the attacker modifies the malicious sample to achieve the effect of deceiving the malware detection classifier deployed on DLaaS. In the testing phase, the classifier identifies carefully modified malware as benign, which can be formalized as follows:\par
\begin{equation}
z^{*}=\arg \min _{z^{\prime} \in Z} \hat{f}\left(\Phi\left(z^{\prime}\right)\right)
=\arg \min _{x^{\prime} \in X} \hat{f}(x^{\prime}),
\end{equation}
where the attacker estimates the target classifier $\hat{f}$ deployed on DLaaS. The Android adversarial sample $z^{\prime}$ extracts the feature vector $x^{\prime}$ through the feature mapping function $\Phi\left (  \right )$ and $z^{*}$ is the successful Android adversarial sample. An evasion attack is essentially the process in which the attacker modifies the features of the malware to minimize the correct output of the classifier.\par
Attackers may have varying degrees of understanding of the target classifier. We mainly discuss the white-box, semi-black-box, and black-box. The prerequisite for the white-box and the semi-black-box is that the attacker has mastered the process of extracting APK features and making selections. In addition, the white-box attack means that the attacker fully understands the classifier deployed on DLaaS and its trained parameters such as the weights and biases. In contrast, in the black box, the attacker does not know how the malware service extracts APK features, and can only upload APK files to DLaaS. Our attack against the malware service is equivalent to a semi-black-box attack. The attacker modifies the APK features and inputs them into the model to obtain the classiﬁer's decisions. Attackers use the DLaaS feedback to improve the understanding of the classifier.\par
\subsection{Attack Restrictions}
When the attacker constructs efficient adversarial samples against the neural network of the malware classifier service, it is very different from the creation of adversarial samples in the image. The key issues are restricted to the following:\par

\begin{enumerate}[]
\item The input domain is changed from continuous and differentiable to discrete. The input is usually a binary vector, where each dimension represents whether the APK has the feature.
\item The visual appearance of the image is required to remain unchanged, but the APK should have the equivalent function after the modification.
\item To modify as few features as possible, attackers need to perform many queries on DLaaS for each feature modified.
\end{enumerate}

Feature deletion: For features from Android manifest.xml, deleting requested permissions affects the function of the application and result in the loss of the permission to access specific resources. The app components declared in the manifest are important parts of Android applications. If app components are easily deleted, the entire application function will be limited. For example, the communication in the Intent component will be affected. Thus, the second restriction will not be met. For features extracted from the classes.dex file, such as suspicious API calls, deleting or encrypting the corresponding API calls  destroys the integrity of the Android application and does not meet the second restriction, so it is very difficult to delete features without damaging the functionality of the Android software.\par
Feature addition: Compared to feature deletion, feature addition is often a safe operation. For features from Android manifest.xml, adding features to the manifest will not affect the execution of existing functions. For the features in the classes.dex file, because we use the Drebin feature extraction method and only a string is extracted, it is safe to add codes that have never been called in classes.dex. To prevent the failure of Android repackaging, we limit the number of added features in subsequent attacks.

\subsection{Craft Android Adversarial Samples}
As mentioned in Section 4.2, an important limitation of crafting Android adversarial samples is to maintain the original malicious functions of the adversarial samples. Therefore, this requires attackers to change as few of the features of the malicious samples as possible. In addition, attackers need to perform many queries on DLaaS to modify a feature, which largely consumes resources. Hence, the best case is that the original sample misleads the classifier by modifying one feature. However, when crafting an Android adversarial sample, the attacker usually needs to modify a few features. The attack method we proposed is different from traditional gradient-based attacks such as FGSM, basic iterative method-a (BIM-a), and basic iterative method-b (BIM-b). This type of gradient attack is to iteratively modify multiple features at a time, and each feature is slightly modified. Due to the first attack restriction described in Section 4.2, the features are discrete, and our method focuses on modifying one feature per iteration. There is no limit to the size of each feature modification, and feature modification can only range from 0 to 1.\par
Crafting Android adversarial samples is formalized as (\ref{con:problem}): in an evade attack, $x=\left ( x_{1},x_{2},\cdots,x_{m}\right )$ is the malicious sample correctly classified by the classifier and $adv$ is the target category of the attack, which is benign. $\delta(x) =\left ( \delta_{1},\delta_{2},\cdots,\delta_{m}\right )$ is the adversarial disturbance determined according to the target category of the attack and the maximum number of features allowed to be modified. We craft adversarial samples based on the following optimization problem to search for the best adversarial disturbance $\delta(x)^{*}$: 
\begin{equation}
\begin{array}{ll}
\underset{\delta (x)^{*}}{\operatorname{maximize}} & f_{adv}(x+\delta (x)) \\
\text { subject to } & \|\delta (x)\|_{0} \leq t,
\end{array} \label{con:problem}
\end{equation}
where $t=1$ is the total number of features that are allowed to be modified in each iteration. The previous gradient attacks often modify multiple features in each iteration. In our proposed attack, only one feature is modified in each iteration, and the modification strength is 1.\par
The usual attack algorithm perturbs all features and cumulatively changes all features, which does not apply to the production of Android adversarial samples. Modifying all the features of the feature vector means modifying the APK file, which often causes great implementation difficulties. Therefore, we use the discrete nature of Android adversarial-sample features to attack a few features without limiting the modification strength.\par
We use the simulated annealing algorithm to craft Android adversarial samples to attack DLaaS. The simulated annealing algorithm is a general probability algorithm, which is used to search for the best adversarial disturbance in an Android feature space within a certain period. Specifically, initialization is required before each iteration, the initial Android sample is the target sample $x_{0}$, and an appropriate value of $T_{0}$ is defined as the initial temperature $T$. The iterative process is the core step of the simulated annealing algorithm, which is divided into the following two parts: 1) the attacker generates new samples locally and 2) queries DLaaS to determine whether it has accepted the new samples.\par
1) Generate new samples: In each iteration, the current sample changes a feature that is zero in the original feature vector to generate a new sample ${x}'$ in the abstract feature space. To avoid falling into the local optimal solution, we use the Tabu list, which stores the changed features. We no longer modify the changed features in the process of generating new samples. In each iteration of generating a new sample ${x}'$, the attacker continuously queries DLaaS to get feedback $f_{adv}({x}')$. We assume that the sample sequences ${x}'_{1},{x}'_{2}\cdots ,{x}'_{t}$ generated by $t$ queries all obey the multidimensional Gaussian distribution. Since a neural network with multiple hidden layers is equivalent to a Gaussian process \cite{neal2012bayesian}, we model the objective function $f_{adv}(\mathbf{{x}'})$ as follows:
\begin{equation}
f_{adv}(\mathbf{{x}'}_{1: t}) \sim \mathcal{G} \mathcal{P}(\mu(\mathbf{{x}'}_{1: t}), k(\mathbf{{x}'}_{1: t}, \mathbf{{x}'}_{1: t})),
\end{equation}
where the mean function is $\mu(\mathbf{{x}'})=0$ and the covariance function k uses the Matern-5/2 kernel. The Gaussian process regression estimates the mean and variance of the true objective function value based on the queried samples.\par
2) Accept new samples: We adopt the upper confidence bounds (UCB) strategy to accept samples\cite{srinivas2009gaussian}. The idea of UCB is to explore samples with large uncertainties, and comprehensively consider the mean and uncertainty of the samples through the Gaussian process that has been modeled. In the $t+1$-th query, we set the objective function to $Fitness\left(\boldsymbol{{x}'_{t+1}} ; {x}'_{1: t}\right)=-(\mu_{t}(\boldsymbol{{x}'_{t+1}})+\sqrt{\beta_{t}} \sigma_{t}(\boldsymbol{{x}'_{t+1}}))$. Through the increment $\Delta Fitness=Fitness({x}')-Fitness(x)$, we can measure whether the new sample is accepted based on the Metropolis criterion. If $\Delta Fitness<0$, accept ${x}'$ as the new current sample $x$; otherwise, accept ${x}'$ as the new current sample $x$ with the probability $\exp (-\Delta Fitness/T)$. If we accept ${x}'_{t+1}$, we use ${x}'_{t+1}$ to query DLaaS and update the Gaussian distribution using (\ref{con:up}). 
\begin{equation}
\begin{aligned}
\mu_{t+1}({x}'_{t+1}) &=k({x}'_{t+1}, {x}'_{1: t}) K_{1: t}^{-1} {f_{adv}}_{1: t} \\
 k_{t+1}({x}'_{t+1}, {x}'_{t+1}) &=-k({x}'_{t+1}, {x}'_{1: t}) K_{1: t}^{-1} k({x}'_{1: t}, {x}'_{t+1})\\
&+k({x}'_{t+1},{x}'_{t+1}),
\end{aligned} \label{con:up}
\end{equation}
where $\mu_{t+1}$ and $k_{t+1}$ are the mean and covariance functions of the posterior distribution on the query of $t+1$ samples, respectively , and $K_{1: t}$ is the covariance matrix of $t × t$. In the iterative query process, the temperature $T$ gradually decreases. When $T$ is low enough or the fitness value reaches the lowest value, we modify the next feature until DLaaS provides incorrect feedback results. The specific algorithm is shown in Algorithm \ref{alg:Framwork}.\par
\begin{algorithm}[h]
\caption{OFEI Attack}
\label{alg:Framwork} 
\hspace*{0.02in} {\bf Input:} \\
\hspace*{0.3in} $x_{0}$: normal malware \textbf{F}:classifier k:maximum number of modified features\\
\hspace*{0.02in} {\bf Output:} \\
\hspace*{0.3in} $x^{*}$: adversarial samples
\begin{algorithmic}[1]
    \State $x^{*} \gets x_{0}$
    \For{$i = 0 \to iteration$}
     \While{$\mathbf{F}\left(x^{*}\right) == 1$ \textbf{and} $\|\delta(x)\|<k$}
                    \State $BestList \gets \{\}$
                    \State $BestFitness \gets \{\}$
                    \State $TabuList \gets \{\}$
                    \State $x \gets getInitSolution(x^{*}, TabuList)$
                    \State $TabuListUpdate()$
                    \State $Best \gets x$
                    \State $Initialize(Fitness(x),T)$
                    \State $InitializeGPmodel(x^{*})$
                    \For{$i = 0 \to MaxQueryTimes$}
                        \State ${x}' \gets getNeighbour(Tabulist)$
                        \State $TabuListUpdate()$
                        \State $\Delta Fitness=Fitness({x}')-Fitness(x)$
                        \If{$\Delta Fitness<0$}
                           \State $ Fitness(x) \gets Fitness({x}')$
                           \State Query and update GPModel with x
                           \State $x \gets {x}'$
                                \If{$Fitness(x)<Fitness(Best)$}
                                    \State $Best \gets x$
                                 \EndIf
                        \Else
                           \State $p=exp (-\Delta Fitness/T)$
                            \If{$p>rand(0,1)$}
                                    \State $Fitness(x) \gets Fitness({x}')$
                                    \State $x \gets {x}'$
                                    \State Query and update GPModel with x
                            \EndIf
                        \EndIf
                       \State $T \gets T\cdot droprate$
                    \EndFor
                   \State $x^{*} \gets x$  
     \EndWhile
       \State $BestList.append(Best)$
       \State $BestListFitness.append(Fitness(Best))$
      \EndFor
       \State {$index \gets \arg \min _{j}BestFitness $}
       \State {$x^{*} \gets BestList[index]$}
     \State \Return{$x^{*}$}
\end{algorithmic}
\end{algorithm}
Due to the randomness of the algorithm, we cannot give a definite execution time complexity, but from qualitative analysis, the algorithm reduces the time complexity of (\ref{con:problem}) from $O(2^{N})$ to $O(iteration*k*MaxQueryTimes)$ in the worst case, which means that the use of the exponential time cost to find the optimal solution is transformed into finding an approximate optimal solution under an acceptable time cost. This algorithm converges to the approximate global optimal solution, which means that the modified feature mostly influences the classification result in each iteration. In each iteration of the query process, the algorithm uses the historical information of the query to construct the posterior distribution of the model, which effectively explores the area of uncertainty and makes full use of the existing information, thereby greatly reducing the number of queries by the attacker. In addition, the algorithm does not require the gradient information of the classifier, so the decision function of the classifier is not required to be differentiable. The attacker only needs to know the probability label, which is a semi-black-box attack scenario. The OFEI algorithm can be combined with other attack algorithms that are not suitable for Android adversarial samples. We will introduce OFEI as an attack framework in the white-box attack scenario in the next section.\par
\subsection{Extend DeepFool and FGSM using OFEI}
Our attack framework can extend the traditional attack methods, so the method of crafting image adversarial samples can be used to create Android adversarial samples. 

DeepFool is essentially based on the Newton fitting method. The adversarial samples are made by continuously making tangents to the given samples to approach the classification hyperplane. This perturbation method is widely used in the image field. DeepFool modifies some pixels of the image in each iteration but this method does not apply to the production of Android adversarial samples for violating the restriction in Section 4.2. Since the features of adversarial samples must be discrete, we can use OFEI as an attack framework to improve DeepFool. We first use the DeepFool method to craft a continuous adversarial sample A. Some of the features of this continuous sample have been changed. We compare A with the original sample to filter out the feature set with the largest change and control the size of this feature set. Finally, we use OFEI to craft adversarial samples under a given feature set which speeds up the convergence.\par

Goodfellow et al. \cite{goodfellow2015explaining} proposed the FGSM method. The main idea of this method is to apply very small linear perturbations to each pixel. When the linear perturbation and the weight of the high-dimensional space are calculated together, it will mislead the classifier. In the production of Android adversarial samples, all features cannot be changed every iteration. Because each feature is discrete, small disturbances are meaningless. However, we can use the direction of each disturbance in the FGSM algorithm to improve the OFEI attack framework. Crafting Android adversarial samples can only add features, so we select features with a positive loss function gradient for modification. In each iteration, the range of features selected for modification is reduced, which speeds up the convergence of the simulated annealing algorithm.

\section{Uncertainties}
In Section 4.3, we propose that the OFEI exploits the uncertainty of the sample to attack DLaaS. Therefore, we use the combination of two uncertainties to detect adversarial samples in this section to provide a more secure DLaaS.\par
Unlike the previous work\cite{ma2018characterizing}\cite{feinman2017detecting}, only one uncertainty was used to measure adversarial samples or adversarial regions. We mainly use two types of uncertainties \cite{kendall2017uncertainties} to construct the combined uncertainty. EU is the uncertainty caused by the model. Because the model does not have enough data to train, the learning is insufficient. AU is the inherent uncertainty of data. We intuitively assume that the process of crafting adversarial samples that is actually a process of changing the two uncertainties of adversarial samples.\par
To capture two uncertainties, we build a Bayesian neural network model \cite{neal2012bayesian}. We put the weights on the Gaussian prior distribution: $W \sim N(0,1)$, and the Bayesian neural network replaces the determined parameters with the distribution of these parameters.\par
EU is represented by the error of the model's parameters. This error reflects the probability of whether sample D is in the distribution that the model has learned. Thus, we exploit the Monte Carlo method to find EU. The Monte Carlo method can produce an estimate of the model's posterior distribution $P(W|D)$ by limited sampling. Through this estimated posterior distribution, we can obtain EU. Specifically, we input the data into the network $T$ times and then take an average of the results to obtain the probability vector $P$ which is the final prediction. We take the entropy of the probability vector $P$ as the EU of the adversarial sample. The calculation of entropy is as (\ref{con:H}):

\begin{equation}
H(p)=-\sum_{c=0}^{1} p_{c} \log p_{c} \label{con:H},
\end{equation}
where c represents the possible categories of the sample, 0 represents that the sample is classified as benign, and 1 represents that the sample is classified as malicious.\par
AU is the inherent error of the data, which is a function of the input data. In malware detection, the error may be caused by feature extraction, which has nothing to do with the amount of data. In AU, we mainly focus on heteroscedastic uncertainty. For different inputs, the errors are different, which means that some samples have more difficulty extracting features and cause different errors. To find heteroscedastic uncertainty, deep learning models can learn to predict uncertainty by using a modified loss function. For classification tasks, the Bayesian deep learning model has two outputs, which not only have the function of predicting the SoftMax value but also obtain the input variance. Predicting the variance of the model is an unsupervised learning process because the model has no variance labels for learning. As shown in (\ref{con:un}):
\begin{equation}
\left[\hat{y}, \hat{\sigma}^{2}\right]=f^{\widehat{W}}(x) \label{con:un},
\end{equation}
where $f$ is a malware classifier improved into a Bayesian neural network and $\widehat{W}$ is the weight of the network. The input of the Bayesian network is the feature vector $x$. Therefore, $\hat{y}$ and $\hat{\delta ^{2}}$ are separately predicted in the network.\par

For the uncertainty of the loss learned in the attenuation process, we use (\ref{con:loss}) to change the loss function. The loss function creates a normal distribution $\quad \epsilon_{t} \sim \mathcal{N}(0, I)$ with a mean of zero and a predicted variance. The predicted logit value is distorted by sampling from the distribution, which is used to calculate the probability value of SoftMax. The loss function obtains Monte Carlo samples by sampling $N$ times and then calculates the classification cross-entropy from the average of the $N$ samples.
\begin{equation}\begin{aligned}
\hat{\mathbf{x}}_{i, n} &=\mathbf{y}_{i}^{\mathbf{W}}+\sigma_{i}^{\mathbf{W}} \epsilon_{n}, \quad \epsilon_{n} \sim \mathcal{N}(0, I) \\
Loss &=-\sum_{i} c \cdot \log \frac{1}{N} \sum_{n} \frac{\exp (\hat{x}_{i, n, c})}{\sum_{c^{\prime}} \exp (\hat{x}_{i, n, c^{\prime}})},
\end{aligned} \label{con:loss}
\end{equation}
where $\mathbf{y}_{i}^{\mathbf{W}}$ and $\sigma_{i}^{\mathbf{W}}$ are the outputs of the Bayesian deep learning model with parameter $\mathbf{W}$, $\hat{\mathbf{x}}_{i, n}$ is the $n$ sampling of the i-th sample, $c$ is the actual classification of the sample, and $c^{\prime}$ is one of the all possible classifications.

\section{Experiments}
In this section, we describe the experimental setup first. Second, we evaluate various attack methods through attack experiments. Finally, we compare the defense effects of various defense methods through defense experiments.
\subsection{Experimental Setup}
\subsubsection{DataSet}
Our experiment is based on the Virusshare\cite{virusshare.com} and Contagio\cite{contagiodump.blogspot.com/} datasets, which contain 23,072 malicious samples. In addition, we collected applications from the Huawei and Xiaomi app stores from January 2020 to February 2021. We exploited the online malware scanning service of VirusTotal to identify the collected applications. When all virus scanners in VirusTotal treat a collected application as a benign application, the application will be included in the set of benign applications. We finally collected 25,855 benign samples.\par
We extracted static features in Drebin for all applications. The features are divided into eight categories, including m features in total. Each feature is a discrete binary value, which represents whether the application has the feature. We directly extracted an Android application into a binary indicator vector used to represent the application. The feature vector is $X \in \left \{0,1\right \}^{m}$, where m = 25,000. The extracted features are shown in TABLE~\ref{tb:table1}.
\subsubsection{Evaluation Metrics for Attack and Defense}
We use the following indicators to measure the efficiency of attacks and defenses: \par
\begin{itemize}
\item Accuracy: During the testing phase, the percentage of Android samples correctly classified by the classifier in the test set.
\item False Negative Rate (FNR): Among the malicious samples, how many malicious samples are predicted to be benign samples.
\item False Positive Rate (FPR): Among all benign samples, how many benign samples are predicted to be malicious samples.
\item Misclassification Rate (MR): The percentage of misclassified malicious samples that suffered an evasion attack. 
\item Average number of perturbations: In a certain DNN, the total number of features modified by the adversarial samples divided by the number of adversarial samples.
\item Average number of queries: In a certain DNN, the total number of queries required to successfully generate adversarial samples divided by the number of successfully generated adversarial samples.
\end{itemize}
\subsubsection{Empirical Setup }
Training and Testing: We divided the dataset into 80\% training data and 20\% test data. In addition, we split the training set into 20\% as the validation set to avoid overfitting during the training process.\par
Malware Detection: Since there is no public DNN network for Android malware detection, we built simple DNN classifiers with different structures and traditional machine learning classifiers on UniCloud \cite{unicloud.com/}. Under the same input features, we chose deep learning architectures with better detection performance than machine learning classifiers. We mainly evaluated the accuracy and FNR of the classifier, which are important indicators to measure the classifier's resistance to evasion attacks. As shown in the TABLE~\ref{tb:table3}, in traditional machine learning methods, we chose random forest, logistic regression, and support vector machine (SVM) methods to detect malware. The results show that the SVM classifier performs best. We chose three deep network architectures that are superior to the SVM algorithm as the attack targets. In all DNN architectures, [200, 100] performs best, [10, 10] has the lowest FNR, and the accuracy and FNR of [200, 200] fall between the two. We used these three networks to study subsequent attacks. In the defense experiment, we uniformly set the original model to [200, 200], to ensure comparability.\par
OFEI attack framework parameter setting: We encoded the Android application into a vector, which is constantly modified by the simulated annealing algorithm. To conduct a full search but not frequently query the DLaaS, we set $\sqrt{\beta_{t}}=0.6$ in fitness. The larger the starting temperature, the easier it is to receive solutions that make the fitness higher, resulting in greater randomness of state jumps. We set the temperature to $T=200$. The cooling speed determines the speed of finding an optimal solution. If the setting is too small, the algorithm will not have time to fully optimize, so we set the temperature drop rate to 0.5. Finally, the adversarial sample with the least disturbances is selected after 10 attacks occur.\par
Attack maximum modification feature settings: We analyzed the features of all software. From TABLE~\ref{tb:table2}, the distance between the first quantile and the third quantile is greater than 20. Therefore, to guarantee that the modified software falls between the first and third quantiles, we set the maximum number of modified features to 20.\par
Other attack method settings: For the pointwise attack, we first continuously enhanced the generation of salt-and-pepper noises and added them to the malicious sample, increasing the noise intensity by 1/1000 each time until the classifier misclassified it as benign or the intensity increased to 1. We repeated the process 10 times and then used the benign samples with minimal disturbance as input to generate adversarial samples by modifying the features. GenAttack is based on a genetic algorithm, which randomly modifies a feature of the original sample to generate an initial population with a population size of 6. Then, the fitness of the population is calculated. If the category of the sample with the highest fitness changes, the adversarial sample is output. Otherwise, the parents are selected according to the fitness. Samples with high fitness are more likely to be selected, while samples with low fitness are eliminated. Crossing the corresponding features of the parents, as long as the corresponding feature is 1, the newly generated feature is set to 1, thereby generating offspring and mutating the offspring with a mutation probability of 5e-2.

\begin{table}
\begin{center}
\small
\caption{Classifier performance for malware detection.}
\begin{tabular}{cccc}
\hline
\textbf{Classifier} & \textbf{Accuracy} & \textbf{FNR} & \textbf{FPR} \\
\hline 
$[10,10]$ & 96.51\% & 1.67\%  & 7.05\%\\
$[10,200]$ & 96.28\% &1.93\%  & 7.70\%\\
$[200,10]$ & 96.35\% & 2.62\%  & 5.93\%\\
$[50,50]$ & 96.30\% & 1.86\%  & 7.81\%\\
$[50,200]$ & 96.33\% & 2.62\%  & 5.98\%\\
$[200,50]$ & 96.34\% & 2.63\%  & 5.93\%\\
$[100,200]$ & 96.37\% & 1.96\%  & 7.35\%\\
$[200,100]$ & 96.63\% & 2.16\%  & 6.04\%\\
$[200,300]$ & 96.37\% & 1.91\%  & 7.47\%\\
$[200,200]$ & 96.53\% & 2.65\%  & 5.36\%\\
$[300,200]$ & 96.33\% & 1.70\%  & 8.04\%\\
$[200,200,200]$ & 96.42\% & 2.47\%  & 6.04\%\\
$[200,200,200,200]$ & 96.44\% & 2.72\%  & 5.41\%\\
RandomForest & 95.56\% & 2.11\%  & 9.64\%\\
LogisticRegression & 96.28\% & 2.34\%  & 6.78\%\\
SVM & 96.48\% &2.16\%  & 5.84\%\\
\hline
\end{tabular}
\label{tb:table3}
\end{center}
\end{table}

\begin{table}
\small
\caption{Feature statistics of Android samples.}
\begin{tabular}{ccccc}
\hline
  & 1st Quantile & Median & 3rd Quantile& Mean \\
\hline 
all  & 29 & 42  & 76 & 87.94\\
malware & 26 & 42  & 49& 47.61\\
\hline
\end{tabular}
\label{tb:table2}
\end{table}

\subsection{Attack Experiment}
\subsubsection{Attack Modes}
We consider the following two attack modes: all features can be modified or only add features in Android manifest.xml.\par
All features: Modify all features, including the features in the code. Since Drebin extracts feature strings, it can be modified by adding pseudocode after returning from the code. However, this attack method is only applicable to the feature extraction method of a specific classifier and is not transplantable.\par
Android manifest.xml only: The feature in the manifest of the Android program is easy to operate. To add features, only adding an extra line to the manifest file is needed and existing functions of the program will not be affected. This attack mode is very effective and has good portability.\par
\subsubsection{Attack Analysis}
In the first attack mode, all features of the Android software can be changed. We applied the attack framework introduced in Section 4 to the malicious samples in the testing phase, including OFEI, OFEI+FGSM, and OFEI+DeepFool. Our attack framework mainly compared attack performance with JSMF, GenAttack, and pointwise attack through the following three aspects: 1) how difficult it is for different attack algorithms to generate adversarial samples; 2) the deceptive effect of adversarial samples on neural networks; and 3) the query efficiency of different attack algorithms on DLaaS.\par
1) The average number of perturbations reflects the difficulty of generating adversarial samples. If the attacker modifies fewer features, then fewer actual changes to the APK occur. The results are shown in TABLE~\ref{tb:table4}. The OFEI attack algorithm causes slightly more disturbances than the JSMF attack algorithm, but they are very close. When the OFEI algorithm is applied to DeepFool or FGSM as an attack framework, the average number of disturbances reduces. The average perturbations of OFEI+FGSM are lower than those of JSMF in the [200, 100], [200, 200] networks. Compared with GenAttack and pointwise attack, the OFEI's disturbances significantly reduce. Among the three network architectures, the [10, 10] network requires the most perturbations, so generating adversarial samples in this network is the most difficult. Conversely, [200, 100] is the easiest for attackers to craft adversarial samples. The results show that the disturbance effect of OFEI as a semi-black-box attack is similar to that of the white-box attack JSMF. OFEI applied to white-box attacks as an attack framework can reduce the average disturbance. The FGSM algorithm using the OFEI attack framework is better than JSMF in most cases. When the OFEI attack framework knows the gradient of the classifier, the number of perturbations reduces, which accelerates the convergence of the OFEI algorithm. In the semi-black-box case, the OFEI attack only modifies one feature through query feedback in each iteration, so the disturbance is fewer than GenAttack based on the genetic algorithm and pointwise attack with salt-and-pepper noise added.\par
\begin{table*}
\begin{center}
\small
\caption{Perturbations of the two attack modes.}
\begin{tabular}{lcccccc}
\hline & \multicolumn{6}{c} { Perturbations } \\
\cline { 2 - 7 }Classifier & OFEI & OFEI+DeepFool & OFEI+FGSM & JSMF & GenAttack & Pointwise\\
\hline $[200,100]$  & 5.64/5.69  & 5.20/5.35 & 4.34/4.40 & 4.40/4.46 & 10.23/10.43  &  17.12/17.25\\
$[200,200]$         & 5.88/5.93  & 5.65/5.82 & 4.57/4.60 & 4.64/4.71 & 10.75/11.20 &  17.35/17.74\\
$[10,10]$           & 6.48/6.54  & 6.14/6.34 & 6.05/6.20  & 5.37/5.42 & 11.22/11.84 &  18.14/18.33\\
\hline
\end{tabular}
\label{tb:table4}
\end{center}
\end{table*}
2) We evaluated the degree that the adversarial sample deceived the classifier through the accuracy and misclassification rate of the classifier. We implemented different attacks on all malicious samples in the test samples. The results are shown in TABLE~\ref{tb:table5}. Under different attacks, the FGSM algorithm using the OFEI attack framework can deceive the most different structures. In the [200,200] network, the misclassification rate of OFEI+FGSM reaches 98.51\%. The misclassification rate of adversarial samples made under the OFEI attack framework is higher than that of JSMF. Since JSMF only considers the forward derivative between the output of the classifier and the input, the OFEI attack framework comprehensively considers the output and the uncertainty of the classifier to modify the features during the attack process, which makes it easier for the generated adversarial samples to fall into the classifier distributions that have not been learned. Compared with GenAttack and pointwise attack, the misclassification rate of OFEI has been greatly improved, because GenAttack and pointwise attack need to modify a large number of features during the attack process, which exceeds the maximum number of modified features and causes the attack to fail.\par
\begin{table*}[h]
\begin{center}
\small
\caption{The attack effect of the two attack modes.}
\begin{tabular}{ccccccc}
\hline
\multirow{2}*{Attack} & \multicolumn{2}{|c|}{[200,100]} & \multicolumn{2}{c|}{[200,200]} & \multicolumn{2}{c}{[10,10]} \\
\cline{2-7} & \multicolumn{1}{|c}{Accuracy} & \multicolumn{1}{c|}{MR} & \multicolumn{1}{|c}{Accuracy} & \multicolumn{1}{c|}{MR} & \multicolumn{1}{|c}{Accuracy} & \multicolumn{1}{c}{MR}\\
\hline
OFEI & 28.40/28.48\% & 97.94/97.38\% & 27.60/27.62\% & 98.25/98.23\% & 28.71/28.77\%            & 97.42/97.33\% \\
OFEI+DeepFool & 28.30/28.33\% & 97.95/97.64\% & 27.20/27.24\% & 98.26/98.25\% & 28.65/28.70\%   &  97.68/97.52\% \\
OFEI+FGSM & 28.19/28.20\% & 98.50/97.72\% & 26.00/26.13\% & 98.51/98.48\% & 28.50/28.64\%       & 97.77/97.27\% \\
JSMF  & 29.00/29.02\% & 97.91/97.12\% & 28.44/28.45\% & 98.21/98.18\% & 29.42/29.73\%           & 97.13/97.20\% \\
GenAttack  & 35.59/35.67\% & 85.16/85.09\% & 34.44/34.57\% & 86.50/86.31\% & 35.66/35.75\% & 85.08/84.87\% \\
Pointwise  & 42.66/42.72\% & 74.95/74.30\% & 41.48/41.61\% & 75.67/76.22\% & 42.84/42.90\% & 74.39/74.09\% \\
\hline
\end{tabular}
\label{tb:table5}
\end{center}
\end{table*}
3) The average number of queries reflects the query efficiency of different attack algorithms on DLaaS. The results are shown in TABLE~\ref{tb:table20}. The OFEI's average number of queries is smaller than that of GenAttack and pointwise attack. This is because OFEI constructs a Gaussian process to predict the next query from a series of sample points in the process of modifying the feature each time, while GenAttack uses genetic algorithms to perform random mutation and crossover based on the current query point, without using the original query information. In addition, the pointwise attack first adds salt-and-pepper noise to generate adversarial samples, which changes many features. When revising the changed features by query, a large number of queries are required to generate the least disturbed adversarial samples.
\begin{table}
\begin{center}
\small
\caption{Queries of the two attack modes.}
\begin{tabular}{lccc}
\hline & \multicolumn{3}{c} { Queries } \\
\cline { 2 - 4 }Classifier & OFEI & GenAttack & Pointwise \\
\hline $[200,100]$ & 1128/1140 & 1559/1575 & 3250/3287 \\
$[200,200]$  & 1176/1188 & 1627/1698 & 3414/3453 \\
$[10,10]$  & 1296/1318 & 1673/1743 & 3633/3681\\
\hline
\end{tabular}
\label{tb:table20}
\end{center}
\end{table}

\subsubsection{Another Attack Mode}
Under the second attack mode, there is a limit to the feature modification range when crafting adversarial samples. We can only add features based on those represented in Android manifest.xml. In this attack mode, the adversarial sample we created can be reconstructed into an APK. Because it will not interfere with any existing functions of the Android application, we only need to add a line statement to the manifest file. With this restriction, the feature range of our adversarial samples is reduced from 25,000 to 24,460.\par
In the attack algorithm of the second attack mode, the optional features of each iteration decrease, which leads to an increase in the number of iterations and slower convergence of the algorithm. As shown in TABLE~\ref{tb:table4} and TABLE~\ref{tb:table20}, compared with the first attack mode, the disturbed features and number of queries to DLaaS of all attack methods increase.\par
As shown in TABLE~\ref{tb:table5}, in the second attack mode, the deceptive effect of the generated adversarial samples on the classifier is similar to the effect achieved in the first attack mode, so the second attack mode will not affect the deceptive effect of the adversarial samples on DNNs.\par
\subsubsection{Change in Fitness Values}
We evaluated the convergence speed of the OFEI algorithm and OFEI as attack frameworks extending different algorithms. The speed of convergence reflects the time complexity of the algorithms. A slow convergence speed indicates that the attacker needs to query a large number of times, causing serious time resource consumption. We measured the convergence properties of the algorithm from two perspectives. First, we measured the convergence speed of the attack algorithm from the change of the fitness value in the process of modifying the feature of the adversarial sample. Second, when modifying each feature, the degree of fitness decline measures how fast the simulated annealing algorithm converges.\par

Specifically, we chose the [200, 200] network. We randomly selected 30 malicious samples from the dataset, used different attacks to craft adversarial samples and observed the convergence process of the fitness values after the malicious sample changed its features.\par

\begin{figure}[htbp]
\centering    
\subfigure[OFEI] 
{
	\begin{minipage}{7cm}
	\centering         
    \includegraphics[width=2.5in, angle=0]{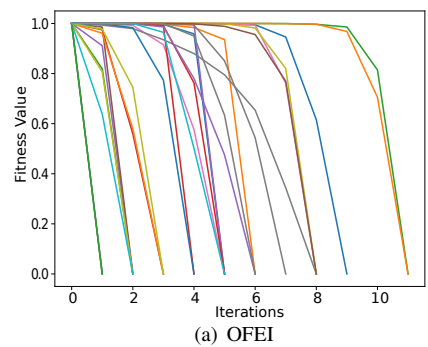}
	\end{minipage}
}
\subfigure[OFEI+DeepFool] 
{
	\begin{minipage}{7cm}
	\centering    
	\includegraphics[width=2.5in, angle=0]{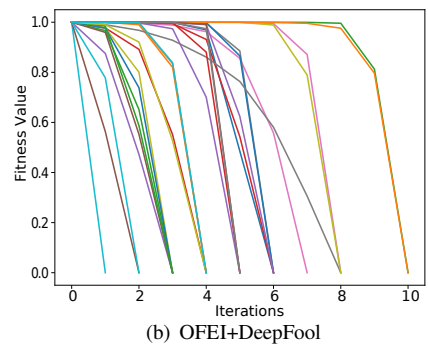}
	\end{minipage}
}
\subfigure[OFEI+FGSM] 
{
	\begin{minipage}{7cm}
	\centering      
	\includegraphics[width=2.5in, angle=0]{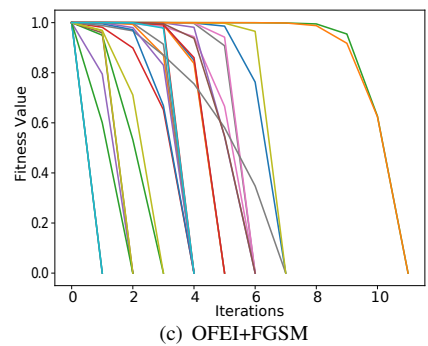}
	\end{minipage}
}
\caption{Changes in the fitness values of 30 randomly selected Android samples during different attacks.} 
\label{fig:fig3} 
\end{figure}

We defined fitness values as the opposite of the probability of a benign category. The purpose of the attack is to minimize the fitness values as much as possible. We compared the fitness convergence of 30 samples of the three attacks in Fig.~\ref{fig:fig3}. OFEI+FGSM converges the fastest, and the fitness value of the malicious sample drops faster during the iteration. The maximum number of OFEI+DeepFool iterations does not exceed 10 features, and OFEI has the slowest convergence.\par
\begin{figure}[htbp]
\centering    
\subfigure[OFEI] 
{
	\begin{minipage}{7cm}
	\centering         
    \includegraphics[width=2.5 in, angle=0]{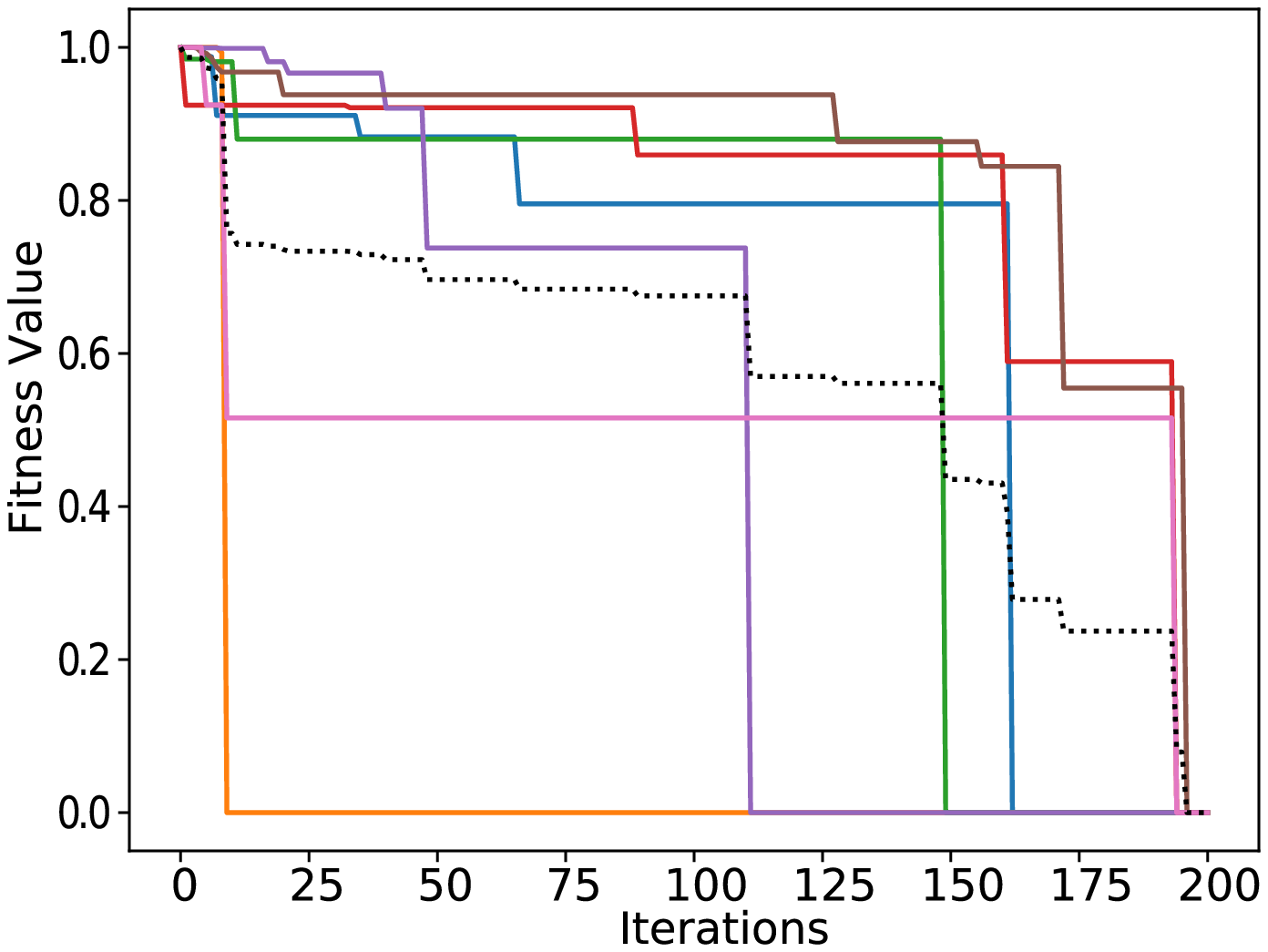}
	\end{minipage}
}
\subfigure[OFEI+DeepFool] 
{
	\begin{minipage}{7cm}
	\centering      
	\includegraphics[width=2.5in, angle=0]{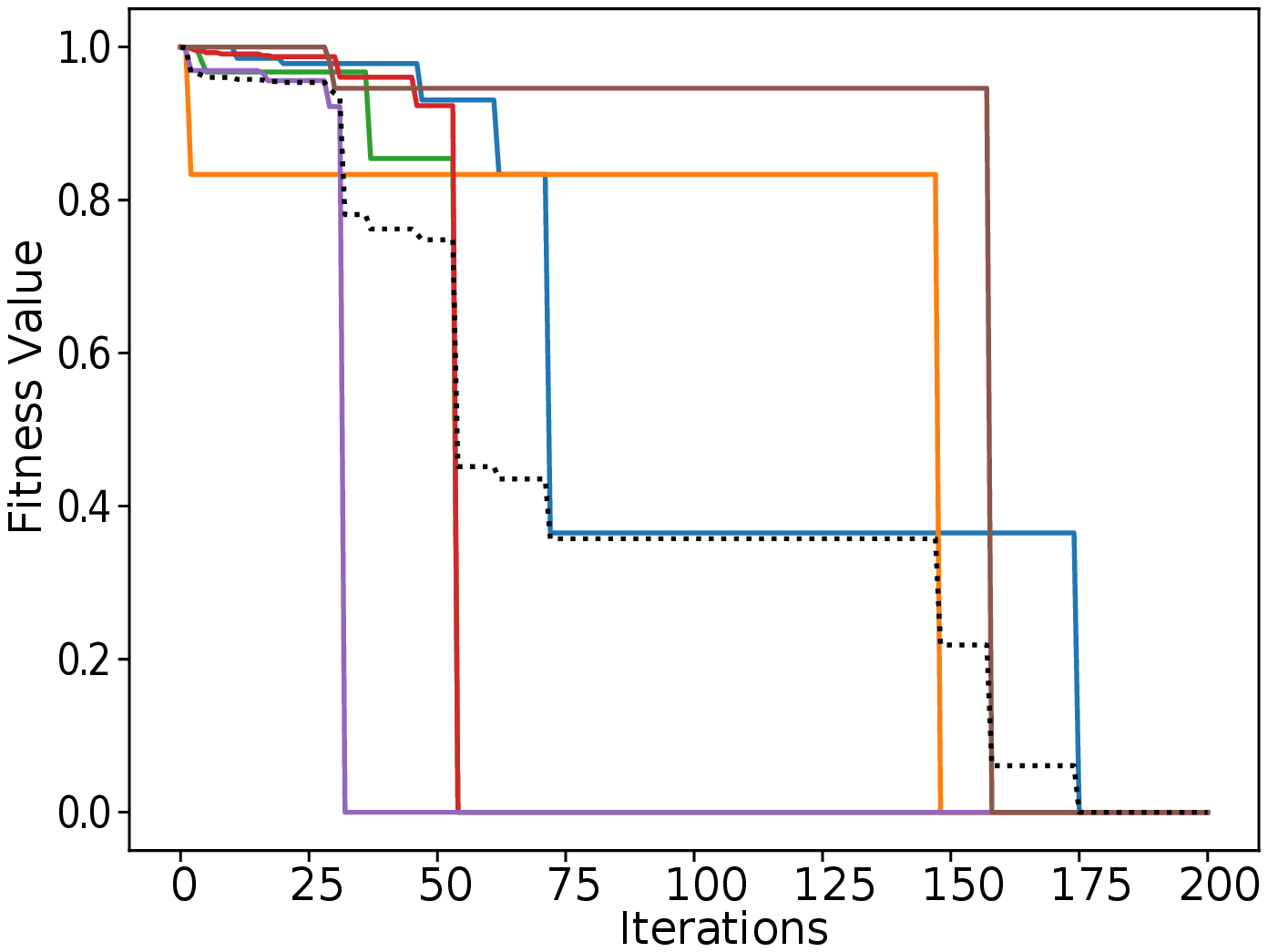}
	\end{minipage}
}
\subfigure[OFEI+FGSM]
{
	\begin{minipage}{7cm}
	\centering     
	\includegraphics[width=2.5 in, angle=0]{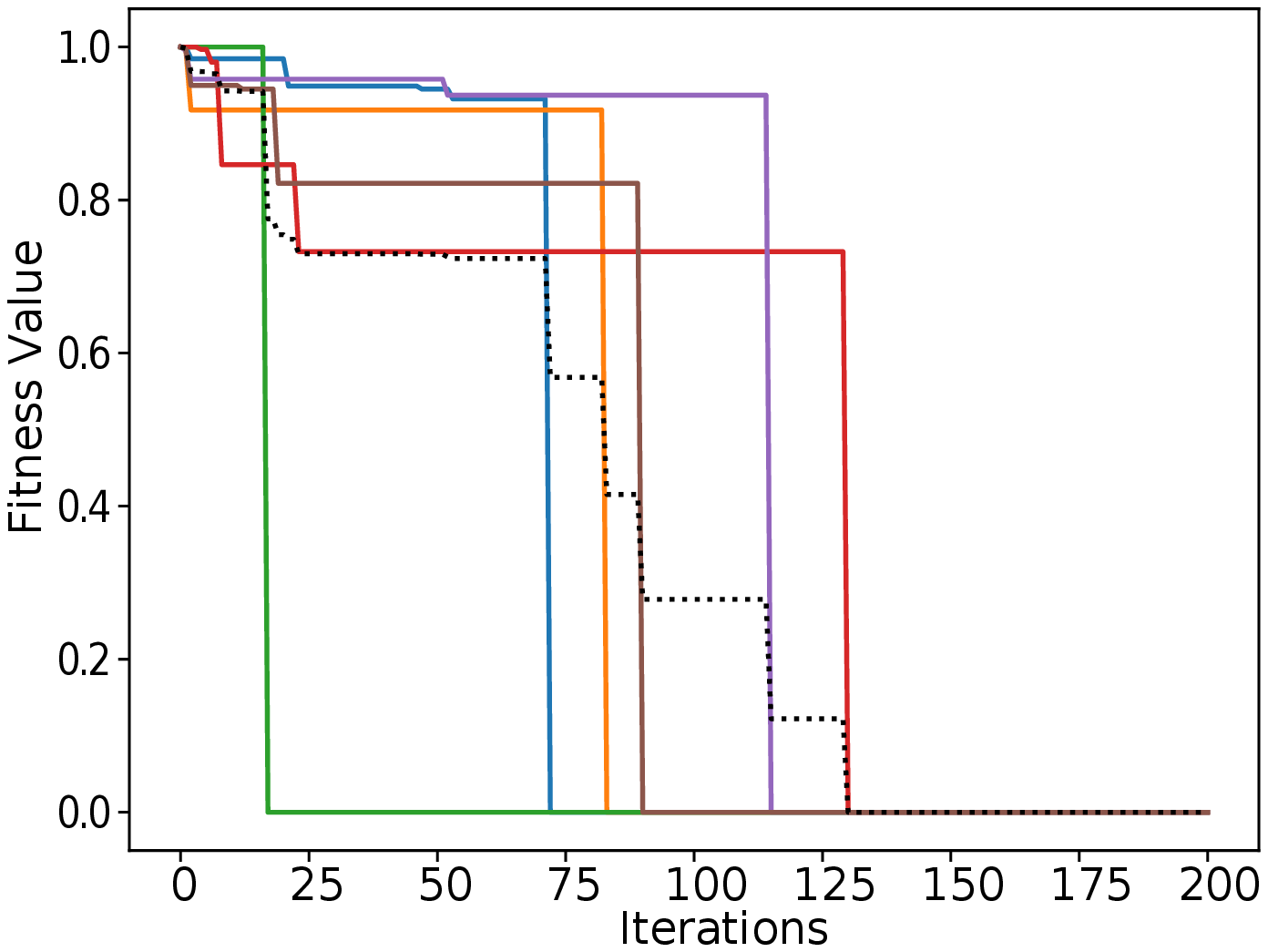}
	\end{minipage}
}
\caption{Simulated annealing process for each feature. The average change of all features is indicated by a dashed line. }
\label{fig:fig4}  
\end{figure}

We randomly took a malicious sample and observed the convergence process of simulated annealing when modifying each feature. The number of iterations means the number of times that DLaaS needs to be queried when modifying a feature. Fig.~\ref{fig:fig4} shows that OFEI generally needs to compare 200 candidate samples when modifying each feature, OFEI+DeepFool generally needs to compare 175 candidate samples and OFEI+FGSM generally needs to compare 130 candidate samples. The simulated annealing algorithm of OFEI+FGSM under each feature has the fastest convergence.

\subsection{Defense Experiment}
In this section, we experiment with different methods of defending adversarial samples. We use the accuracy and misclassification rate introduced in Section 6.1.3 as indicators. By comparing the model imposing defense methods with the original model, we concluded the impact of different defense methods on the indicators. 
\subsubsection{Characteristics Analysis}
We chose three types of features, i.e., kernel density (KD), local intrinsic dimensionality (LID), and combined uncertainty (CU) to measure the Android adversarial samples. Among them, we subdivided Bayesian uncertainty into EU and AU, where $CU=EU+\lambda AU$, and we set $\lambda$ to 0.4.\par
\begin{table*}
\begin{center}
\small
\caption{Characteristics analysis of adversarial samples.}
\begin{tabular}{lccccc}
\hline
\multirow{2}*{Attack}& \multirow{2}*{$\frac{ \mathrm{KD}(x^{*})}{\mathrm{KD}(x)}<1$}& \multirow{2}*{$\frac{ \mathrm{AU}(x^{*})}{\mathrm{AU}(x)}>1$}& \multirow{2}*{$\frac{ \mathrm{EU}(x^{*})}{\mathrm{EU}(x)}>1$} & \multirow{2}*{$\frac{ \mathrm{CU}(x^{*})}{\mathrm{CU}(x)}>1$}& \multirow{2}*{$\frac{ \mathrm{LID}(x^{*})}{\mathrm{LID}(x)}>1$} \\
& & & & &\\
\hline 
OFEI & 90.97\% & 88.98\% & 94.72\% & 97.39\% & 87.88\% \\
JSMF & 89.71\% & 85.33\% & 93.75\% & 96.52\% & 87.30\% \\
GenAttack & 94.45\% & 92.56\% & 86.24\% & 96.42\% & 93.21\% \\
Pointwise  & 94.05\% & 91.24\% & 85.65\% & 96.25\% & 94.47\%\\
\hline
\end{tabular}
\label{tb:table9}
\end{center}
\end{table*}
As shown in TABLE~\ref{tb:table9}, when we generate Android adversarial samples with different attack methods, EU, AU, CU and LID are usually larger than normal samples while KD is usually smaller than in the normal samples. This makes sense. Because the adversarial sample is artificially and unnaturally constructed. The probability density of the adversarial space is low, so the kernel density is low. From the perspective of the manifold, the adversarial space spans multiple manifolds and has a more complex local structure, so the adversarial samples located in the adversarial space have higher LID values. From the perspective of uncertainty, the reason for the adversarial deception model is that the learned distribution of the model is deceived. Therefore, the higher EU of the adversarial sample means that the model parameter error caused by the adversarial sample is larger, which causes the incorrect classification result to evade attack. AU is the inherent nature of the data itself. The perturbation added by the attacker to generate adversarial samples is equivalent to adding noise to the data itself and misleads the classifier by changing the nature of the data itself. So the adversarial sample has higher AU. We construct the CU, which combines the properties of two uncertainties to measure the adversarial samples. Bayesian uncertainty not only considers the impact of the adversarial sample on the model, but also considers the nature of the input space where the adversarial sample is located. We observe the Android adversarial samples generated by different attacks. KD, CU, and LID are all effectively distinguished. This also proves the versatility of these three types of features.

\subsubsection{Detector}
Our purpose of training the detector is to better understand the area where the adversarial sample is located. We chose a simple logistic regression model as the classifier. This classifier was used to distinguish adversarial samples from normal samples. KD, LID, AU, EU and CU was respectively used as the input feature of the classifier. We used the characteristics of adversarial samples to train the classifier, where adversarial samples came from OFEI or JSMF, and observed the detection efficiency of the classifier against different attacks. The results are shown in TABLE~\ref{tb:table10}. We compared the accuracy of the five classifiers. CU improves the detector compared to AU and EU. AU has a good detection effect on GenAttack and pointwise attack, because these two attacks both impose noise or feature cross-mutation during the attack process, so AU can be distinguished by the nature of the data itself. EU can be used to well distinguish OFEI because OFEI perturbs uncertain features to deceive the classifier during the attack. CU combines AU and EU, so the detection effect is better than that of KD and LID. Both KD and LID use the surrounding samples of the adversarial sample to measure the adversarial space. Therefore, it is not effective to distinguish between JSMF and OFEI attacks through small disturbances. In summary, we used CU as the feature of the detector in the subsequent experiments.  
\begin{table}

\small
\caption{Performance of detecting adversarial samples.}
\begin{tabular}{p{0.6cm}<{\centering}|p{0.5cm}<{\centering}|p{0.9cm}<{\centering} p{0.9cm}<{\centering} p{2cm}<{\centering} p{1.5cm}<{\centering}}
\hline 
Train & Test & OFEI & JSMF & GenAttack & Pointwise \\
\hline 
\multirow{5}*{JSMF}& KD & 92.72\% & 93.50\% & 96.69\% & 96.30\%\\
 & LID & 90.03\% & 91.93\% & 96.70\% & 95.89\% \\
& AU & 92.23\% & 88.33\% & 95.16\% & 95.07\% \\
& EU & 96.86\% & 96.51\% & 90.41\% & 91.48\%  \\
& CU & $\mathbf{98.55\%}$ & $\mathbf{98.03\%}$ & $\mathbf{99.28\%}$ & $\mathbf{98.15\%}$  \\
\hline 
\multirow{5}*{OFEI}& KD & 92.42\% & 93.29\% & 96.81\% & 96.36\% \\
 & LID & 90.75\% & 91.20\% & 96.58\% & 95.74\%  \\
& AU & 92.06\% & 88.61\% & 95.45\% & 95.37\% \\
& EU & 96.58\% & 96.44\% & 90.68\% & 91.53\% \\
& CU & $\mathbf{98.53\%}$ & $\mathbf{98.28\%}$ & $\mathbf{97.83\%}$ & $\mathbf{98.26\%}$ \\
\hline
\end{tabular}
\label{tb:table10}
\end{table}

\subsubsection{Comparisons With Other Defensive Methods}

To measure the performance of our proposed detector using uncertainty training, we compared our defense method with other traditional defense solutions. We chose a deep learning architecture of [200, 200]. Comparing the defense method with the original model, we observed changes in various indicators.\par

\begin{figure*}[htbp]
\centering
\subfigure[Accuracy]
{\includegraphics[width=3.5in, angle=0]{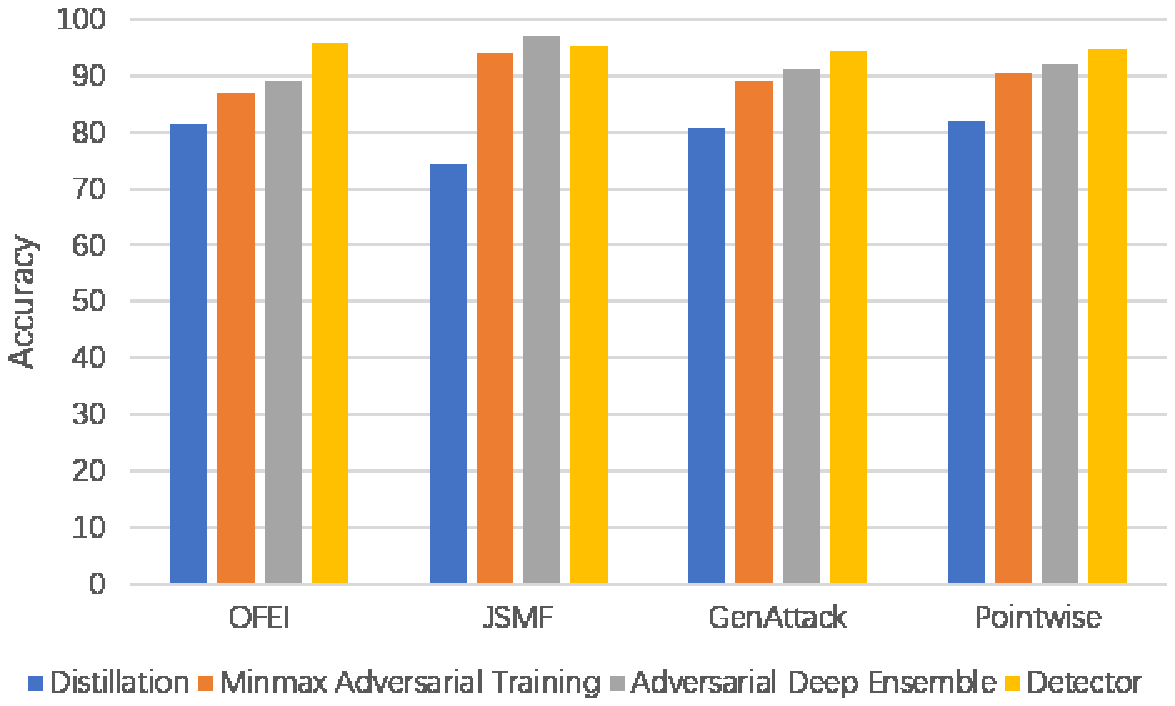}}
\subfigure[Misclassification rate]
{\includegraphics[width=3.5in, angle=0]{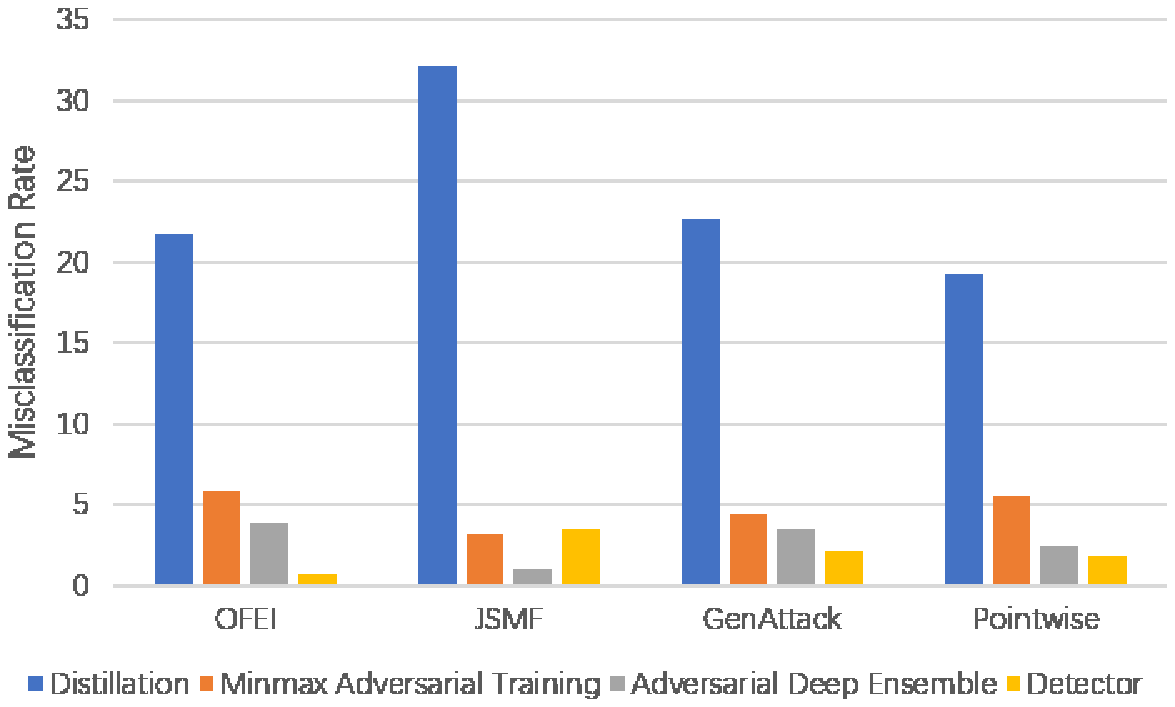}}
\caption{Comparison of different defense methods.}
\label{fig:fig5}                
\end{figure*}

First, we made comparisons using the distillation method \cite{papernot2016distillation}. We set the distillation temperature to 150, trained a network [200, 200] to obtain the original model on the training set $X$, and then used the probability label of the original model to train a distillation model on the same training set and temperature for defense. In the first attack mode, we replaced the malicious samples with adversarial samples made by different attack methods during the testing phase. We observed the defense effect of the distillation model compared to the original model. Fig.~\ref{fig:fig5} shows that the misclassification rate significantly drops. In the original model [200, 200], the misclassification rate reaches more than 98\%, but after using distillation, the misclassification rate drops to approximately 30\%. The accuracy of the classifier is also greatly improved, from 26\% to approximately 80\%. Although the defense effect is significantly improved, the misclassification rate is still high.\par
Second, we used minmax adversarial training \cite{al2018adversarial} for the network [200, 200]. We used JSMF to generate adversarial samples for all malicious samples in the training data, and then minimized the loss of adversarial samples and benign samples during the training process to enhance the robustness of the model. As shown in Fig.~\ref{fig:fig5}, the model trained with JSMF adversarial samples is more effective for the defense of JSMF adversarial samples, and the misclassification rate of other adversarial samples is also significantly improved. Minmax adversarial training assumes that the model already has the loss value of the adversarial sample as prior knowledge. However, the OFEI does not depend on gradient information, so minmax adversarial training is not good for its defense effect. In addition, we used an adversarial deep ensemble \cite{li2020adversarial} to enhance the robustness in the ensemble network [200, 200], [200, 100], [10, 10]. We performed minmax adversarial training for each base classifier, and then merged them. Compared with the minmax adversarial training model, the robustness is improved, but the defense against attacks other than JSMF is still slightly worse.\par
Finally, we used the combination of uncertainties to detect adversarial samples. When training the detector, we selected the adversarial samples generated by the JSMF attack to train the detector. The detector first judged whether the sample was adversarial. If it was an adversarial sample, it was classified as malware, and if it was not an adversarial sample, it was input to the neural network detection software with a subsequent architecture of [200, 200]. We observed the defensive effect of the detector model on different attacks. From the Fig.~\ref{fig:fig5}, the accuracy of the detection model is higher than that of the model using minmax adversarial training. The misclassification rate of the detection model greatly decreases and approaches zero. The detector comprehensively considers the impact of the adversarial sample on the model and the difference between the adversarial sample itself and the normal sample for defense. Therefore, OFEI, which uses uncertainty to interfere with features, can be effectively defended. Although the traditional defense method can increase the generalization of the model, it cannot distinguish the characteristics of the adversarial sample itself. Therefore, the traditional defense method is more coarser-grained than the detector using uncertainty.\par We compared the three defense methods as follows. If enough adversarial samples are added, the accuracy of the traditional adversarial training model will be greatly improved. However, a large number of adversarial samples need to be produced. In contrast, the detector model we use does not need to know the adversarial samples in advance to achieve high defense efficiency, and the defense effect is the best of the three methods.

\section{Conclusion and Future Work}
\label{section:Conclusion}
In this paper, we present a novel OFEI attack framework to craft Android adversarial samples for malware classifier services that offer label information from DLaaS. The experimental results show that the performances of our attack method are better than GenAttack and pointwise attack on the Virusshare and Contagio datasets. In addition, our proposed attack framework can extend DeepFool and FGSM to fit the limitations of Android adversarial sample production. OFEI+DeepFool reaches the attack effect similar to JSMF after improvement and OFEI+FGSM performs better than that of JSMF. 

Furthermore, to provide a more secure malware classifier service, we mainly use the combined uncertainty to identify Android adversarial samples. We compare the uncertainty-based detector with the traditional defense mechanisms. Although minmax adversarial training and adversarial deep ensembles can significantly reduce the misclassification rate of the classifier, a large number of adversarial samples are required. When facing other attack methods, the adversarial training effect becomes worse and lacks portability. For the uncertainty-based detector, we use it to detect and exclude the adversarial samples before inputting them into the subsequent classifiers. The experimental results shows that the detector model has achieved good defense effects when faced with different attacks.\par
\bibliographystyle{IEEEtran}
\bibliography{main}


\end{document}